\documentclass[twocolumn,english,aps,prd,reprint,floatfix,notitlepage,footinbib,preprintnumbers,superscriptaddress]{revtex4-1}
\pdfoutput=1
\usepackage{lmodern}

\usepackage[T1]{fontenc}
\usepackage[latin9]{inputenc}
\usepackage{geometry}
\usepackage{subfigure,lmodern, amsmath,amssymb, graphicx, pifont, adjustbox, bm, xcolor}
\usepackage{amsfonts}
\usepackage{geometry}
\usepackage{comment}
\usepackage{mathtools}
\usepackage{float}
\usepackage{slashed}
\usepackage{ragged2e}
\usepackage{array}

\makeatletter\g@addto@macro\bfseries{\boldmath}\makeatother

\newcommand{\SigmaScaled}{\scalebox{1.7}{\raisebox{-0.4mm}{$\Sigma$}}}

\usepackage{stackengine}
\usepackage{esint}
\usepackage[unicode=true,pdfusetitle,
 bookmarks=true,bookmarksnumbered=false,bookmarksopen=false,
 breaklinks=false,pdfborder={0 0 1},backref=false,colorlinks=true]
 {hyperref}
\hypersetup{
 pdfauthor={Clifford Cheung and Grant N. Remmen},
 citecolor=black,linkcolor=black,urlcolor=black}

\newcommand{\appendixref}[1]{\hyperref[#1]{appendix~\ref{#1}}}
\def\equationautorefname~#1\null{eq.\,(#1)\null}
\usepackage{breakurl}
\usepackage{breakurl}
\usepackage[hang,flushmargin]{footmisc} 
\allowdisplaybreaks
\makeatletter

\usepackage{etoolbox}
\apptocmd{\thebibliography}{\justifying\setlength{\leftskip}{7.4mm}}{}{} 
 
 \usepackage{relsize}
\usepackage{babel}

\makeatletter
\def\simgt{\mathrel{\lower2.5pt\vbox{\lineskip=0pt\baselineskip=0pt
           \hbox{$>$}\hbox{$\sim$}}}}
\def\simlt{\mathrel{\lower2.5pt\vbox{\lineskip=0pt\baselineskip=0pt
           \hbox{$<$}\hbox{$\sim$}}}}
\makeatother

\usepackage{changepage}

\newcommand{\be}{\begin{equation}}
\newcommand{\ee}{\end{equation}}
\newcommand{\bea}{\begin{eqnarray}}
\newcommand{\eea}{\end{eqnarray}}

\newcommand{\Eq}[1]{Eq.~(\ref{#1})}

\newcommand{\eq}[2]{\be\begin{aligned}#1 \label{#2}\end{aligned}\ee}

\newcolumntype{P}[1]{>{\centering\arraybackslash}p{#1}}

\begin{document}

\title{Stringy Dynamics from an Amplitudes Bootstrap}

\author{Clifford Cheung}
\affiliation{Walter Burke Institute for Theoretical Physics, California Institute of Technology, Pasadena, CA 91125}
\author{Grant N. Remmen}
\affiliation{Kavli Institute for Theoretical Physics and Department of Physics, University of California, Santa Barbara, CA 93106}
    
\begin{abstract}
\noindent 
We describe an analytic procedure whereby scattering amplitudes are bootstrapped directly from an input mass spectrum and a handful of physical constraints: crossing symmetry, boundedness at high energies, and finiteness of exchanged spins.
For an integer spectrum, this procedure gives a first principles derivation of a new infinite parameter generalization of the Veneziano amplitude that is unitary while exhibiting dual resonance and consistent high-energy behavior.
Lifting to a $q$-deformed integer spectrum, we derive the Coon amplitude and its analogous generalizations.  
Finally, we apply this logic to derive an infinite class of deformed Virasoro-Shapiro amplitudes.

\end{abstract}
\maketitle

\preprint{CALT-TH 2023-006}

\maketitle

\noindent {\bf Introduction.} String theory is our leading candidate for a unified and self-consistent formulation of quantum gravity.  Its core assertion is exceedingly simple: perhaps the inventory of the universe is composed not of point particles, but rather of extended objects.  This line of inquiry has sparked an extraordinary influx of new ideas across disciplines, including quantum field theory, phenomenology, cosmology, condensed matter theory, quantum information theory, and even mathematics. 

Despite these successes, pressing questions remain.  How unique is string theory?  What is the minimal set of principles from which it follows?  And most importantly, is it the only solution satisfying these criteria?

The amplitudes bootstrap is a versatile framework for interrogating these questions with mathematical precision.  In this approach, one sculpts out a consistent space of theories by constraining an {\it ansatz} scattering amplitude with certain physical conditions.  For example, imposing Lorentz invariance and factorization on the scattering of massless particles of spin one or two is sufficient to  {\it uniquely} fix the dynamics, thus deriving gauge theory and gravity without the aid of an action~\cite{Elvang:2015rqa,Cheung:2017pzi,Arkani-Hamed:2017jhn,Benincasa:2007xk,Cohen:2010mi}.

In this paper, we show that this approach is even more powerful if we {\it also assume a spectrum} together with a short list of physical constraints that include crossing symmetry, bounded high-energy scaling, and finite spin for the exchanged states.  Imposing these conditions, we derive a new infinite parameter class of amplitudes that subsumes the renowned Veneziano~\cite{Veneziano:1968yb} and Coon~\cite{Coon} amplitudes as special cases.  We then use analogous logic to derive generalizations of the Virasoro-Shapiro amplitude~\cite{Virasoro,Shapiro}, which describes gravitational scattering.

\medskip

\noindent {\bf Physical Constraints.}  We construct four-point, tree-level scattering amplitudes for massless scalars subject to a handful of simple physical criteria:

\medskip
\noindent {\it i) Crossing Symmetry.} For external scalars exhibiting cyclic or full permutation invariance, the corresponding scattering amplitudes satisfy $A(s,t)=A(t,s)$ and $M(s,t)=M(t,s)=M(s,u)=\cdots$, respectively.  Here $A(s,t)$ and $M(s,t)$ can be viewed as gauge theory and gravity amplitudes stripped of their kinematic prefactors, ${\cal F}^4/u$ and ${\cal R}^4$, which are products of the linearized field strengths and curvatures encoding polarizations.  For the rest of this section we focus on $A(s,t)$, though our discussion generalizes straightforwardly to $M(s,t)$.

\medskip

\noindent {\it ii) Polynomial Residues.}  Exchanged states exhibit a finite tower of spins~\footnote{In contrast, locality-violating infinite spin exchanges have been investigated recently in the context of effective field theory bounds~\cite{Caron-Huot:2020cmc} and graviton scattering~\cite{Huang:2022mdb}.}.  Concretely, for a mass spectrum defined by $m^2_n$ for nonnegative integer $n$, the residue on each pole in $s$ is a polynomial in $t$ of degree $n$,
\eq{
R_n(t) = \underset{s=m^2_n} {\!\!\textrm{Res}}A(s,t) =  \sum_{m=0}^n \lambda_{n,m} t^m ,
}{residue_ansatz_linear}
encoding exchanges up to spin $n$. An analysis of more general polynomials, such as those in Ref.~\cite{Cheung:2022mkw}, will be left for future work.

\medskip

\noindent {\it iii) High-Energy Boundedness.} The amplitude vanishes in the high-energy Regge limit defined by sending $s\rightarrow \infty$ at a fixed value of $t$ chosen to be below the gap.  In this case we can derive the 
dispersion relation, 
\eq{
 A(s,t)  = \oint\limits_{s'{=}s} \frac{{\rm d}s'}{2\pi i} \frac{A(s',t)}{s'-s} 
=\int\limits^\infty_{-\infty} \frac{{\rm d}s'}{\pi}  \frac{\textrm{Im} \, A(s',t)}{s'-s} \,,
}{} 
since $A(s,t)$ has poles neither in the $u$ channel nor at infinity~\footnote{The assumption that $A(s,t)<1$ in the Regge limit implies that the polarization-dressed gauge theory amplitude is bounded by ${\cal F}^4 A(s,t)/u < s$.  Remarkably, in all known ultraviolet complete examples, $M(s,t)<s^{-2}$ is superconvergent, in which case the polarization-dressed gravitational amplitude is Regge bounded by ${\cal R}^4 M(s,t)< s^2$.  While our bootstrap procedure only assumes $M(s,t)<1$, superconvergence nevertheless emerges automatically.}.  At tree level, one has $\textrm{Im} \, A(s',t) = \pi \sum_{n=0}^\infty \delta(s'\,{-}\,m^2_n) R_n(t)$, so $A(s,t)$ admits a dual resonant representation as a sum over $s$-channel poles,
\eq{
A(s,t) &= \sum_{n=0}^\infty \frac{R_n(t)}{m^2_n-s}\, ,
}{amplitude_ansatz_cyclic}
which converges for $t <m_0^2$ or when evaluating residues on poles in
$s$ at generic $t$.  Hence, dual resonance is an {\it automatic byproduct} of the vanishing boundary term.  In string theory this happens because the worldsheet can be deformed to exhibit exchanges solely in a single channel.

The last and most nontrivial ingredient in our analysis is the {\it spectrum} of the theory, input via some sequence $m^2_n$.  We will consider integer and $q$-integer spectra, leaving more general possibilities for future work.

To implement the bootstrap described above, we assume a dual resonant ansatz for an amplitude with polynomial residues.  This leaves crossing symmetry as the final and most difficult condition to impose.  As we will see, one can reduce this complex analysis problem in two variables to a simpler problem of a single variable by restricting $t$ to a wisely chosen function of $s$.

\bigskip

\noindent {\bf Integer Spectrum Bootstrap.} To begin, let us consider an integer mass spectrum,
\eq{
m^2_n = n \,,
}{spectrum_linear}
 in which case the dual resonant form of the amplitude is
\eq{
A(s,t) &= \sum_{n=0}^\infty \frac{R_n(t)}{n-s}\, ,
}{amplitude_ansatz_linear}
where $R_n(t)$ is defined in \Eq{residue_ansatz_linear}.  Here we again note that \Eq{amplitude_ansatz_linear} only converges for $t<0$ or on poles in $s$.

Next, we study a special kinematic regime in which $t$ is displaced from $s$ by a positive integer $k$,
\eq{
t=s-k \, .
}{t(s)_linear}
By inserting \Eq{t(s)_linear} into \Eq{amplitude_ansatz_linear} and imposing crossing symmetry, we find that $A(s,s\,{-}\,k)=A(s\,{-}\,k,s)$.  This implies that $\sum_{n=0}^\infty \tfrac{R_n(s-k)}{n-s}\,{=}\, \sum_{n=0}^\infty \tfrac{R_n(s)}{n+k-s}$, which upon a relabeling of terms becomes
\eq{
 \sum_{n=k}^\infty \frac{R_{n}(s-k)-R_{n-k}(s)}{n-s} =-\sum_{n=0}^{k-1} \frac{R_{n}(s-k)}{n-s} \,.
}{diff_amplitude_linear}
The right-hand side comprises a finite collection of terms whose poles are at $s=n<k$.  Since this expression is elsewhere regular, it has vanishing residues at $n\geq k$.  Demanding the same of the left-hand side yields
\eq{
R_n(n-k) = R_{n-k}(n)  
}{residue_crossing_linear}
for $ 1\leq k\leq n$.
Note that \Eq{residue_crossing_linear} should be handled with care since it is neither necessary nor sufficient for enforcing crossing symmetry for all $s$ and $t$.  
It is not sufficient because it was derived for the special kinematic choice in \Eq{t(s)_linear}, so crossing might hold on the support of that condition but not away from it.  Meanwhile, it is not necessary because it originates from the infinite sum in Eq.~\eqref{amplitude_ansatz_linear}, whose residues converge at poles in $s$ for generic $t$ but not necessarily for $t$ satisfying \Eq{t(s)_linear}.
Nevertheless, for the present analysis we take \Eq{residue_crossing_linear} as an input assumption.  Crucially, in all of our later examples, we find that the sums in Eqs.~\eqref{amplitude_ansatz_linear} and \eqref{diff_amplitude_linear} indeed converge and can be trusted. Our resulting amplitudes also exhibit full crossing symmetry for generic $s$ and $t$.

The residue crossing condition in \Eq{residue_crossing_linear}  imposes $n$ constraints on the $n\,{+}\,1$ free parameters in \Eq{residue_ansatz_linear}.  Concretely, we eliminate $\lambda_{n,m}$ for $m\;{<}\;n$ in terms of $\lambda_m \equiv \lambda_{m,m}$, yielding the general solution 
\eq{
R_n(t) 
&= \sum_{m=0}^{n}\frac{ \lambda_{m}}{m!}  \frac{t!}{(t-m)!}\frac{n!}{(n-m)!} \, ,
}{residue_solution_linear}
where we analytically continue $x!$ to $\Gamma(x\,{+}\,1)$ at will. Physically, $\lambda_m$ controls the exchange of spin $m$ modes at levels $n\geq m$.
While \Eq{residue_solution_linear} naively admits arbitrary $\lambda_m$, many choices render the sum in \Eq{amplitude_ansatz_linear} nonconvergent.  For example, this happens if we choose $\lambda_m= \delta_{mm'}$ for some $m'$.  Consequently, it is necessary to determine $\lambda_m$ for which the expression in \Eq{amplitude_ansatz_linear} actually converges.

\medskip

\noindent {\bf Veneziano Amplitude.} Evaluating \Eq{residue_solution_linear} for
 the very special choice $\lambda_m = \tfrac{1}{m!}$,
the Vandermonde identity implies that $
R_n(t) = \frac{(t+n)!}{t! n!} = \begin{psmallmatrix} t+n \\ n \end{psmallmatrix}$.
Inserting this residue into the dual resonant ansatz in \Eq{amplitude_ansatz_linear} yields
\eq{
A(s,t) =\sum_{n{=}0}^\infty \frac{1}{n-s}   \left(
\begin{array}{c}
t +n \\
n
\end{array}
\right)= \frac{\Gamma(-s)\Gamma(-t)}{\Gamma(-s-t)} \, ,
}{eq:Veneziano}
which is the Veneziano amplitude.
Here we obtained right-hand side by relating the sum to a hypergeometric function and applying the Gauss summation theorem.  

\medskip

\noindent {\bf Hypergeometric Amplitude.}
Next, let us consider the more general parameter choice $\lambda_m =\tfrac{r!}{(m+r)!}$
for any {\it real} value of $r$. 
Here the ansatz residue in \Eq{residue_solution_linear} becomes $R_n(t) = \frac{(t+n+r)! r!}{(t+r)! (n+r)!}$,
so \Eq{amplitude_ansatz_linear} evaluates to
\eq{
A(s,t) &=  \sum_{n=0}^\infty \frac{1}{n-s}  \frac{(t+n+r)! r!}{(t+r)! (n+r)!} \\
&= - \frac{1}{s}  {}_3  F_2\left[
\begin{array}{c}
1,-s,1+t+r\\
1-s,1+r
\end{array}
;1\right]  .
}{amplitude_hypergeo_pre}
Via a Thomae transformation~\cite{Bailey}, 
this expression equals
\eq{
A(s,t) = \frac{\Gamma(-s)\Gamma(-t)}{\Gamma(-s-t)}  {}_3  F_2\left[
\begin{array}{c}
-s,\,-t,\, r\\
-s-t,\, 1+r
\end{array}
;1\right] ,
}{amplitude_hypergeo}
which is the Veneziano amplitude times a manifestly crossing symmetric function of $s$ and $t$ that reduces to $1$ at $r=0$.  For arbitrary kinematics, \Eq{amplitude_hypergeo} will be our definition of a new hypergeometric amplitude.

We can recast \Eq{amplitude_hypergeo} in several illuminating forms.  For nonzero real $r$, inserting the definition of the hypergeometric function yields an infinite sum
 over a special case of the generalized amplitudes derived in Ref.~\cite{Cheung:2022mkw},
\eq{
A(s,t) &= \sum_{n=0}^\infty \frac{1}{n!} \frac{r}{r+n}\frac{\Gamma(-s+n)\Gamma(-t+n)}{\Gamma(-s-t+n)} \, ,
}{}
which also coincides with a particular choice of coefficients in the string-inspired ansatz in Ref.~\cite{Gross:1969db}. 

For nonnegative integer $r$,  \Eq{amplitude_hypergeo} is also equal to
\eq{
 A(s,t) {=} ({-}1)^r r! \frac{\Gamma({-}s{-}r)\Gamma({-}t{-}r)}{\Gamma({-}s{-}t{-}r)} + \frac{P_{2r{-}2}(s,t)}{Q_{2r}(s,t)} \,,
}{amplitude_hypergeo_simp}
where $P_{2r-2}$ and $Q_{2r}(s,t) = \prod_{\ell=1}^r (s+\ell)(t+\ell)$ are polynomials of degree $2r-2$ and $2r$, respectively. 
Here the first and second terms in \Eq{amplitude_hypergeo_simp} exhibit spurious poles in $s$ and $t$ that precisely cancel.  Note that \Eq{amplitude_hypergeo_simp} can be analytically continued to negative integer $r$ if we also divide $\lambda_m$ and thus $A(s,t)$ by an overall factor of $r!$.  In this case the hypergeometric amplitude $A(s,t)$ is given just by the first term in \Eq{amplitude_hypergeo_simp}, which is also the basis amplitude comprising the string ansatz in Ref.~\cite{Gross:1969db}.

Next, it will be instructive to study various limits of $A(s,t)$.  For example, for nonnegative integer $r$, the low-energy expansion of the amplitude is
\eq{
& A(s,t) = -\tfrac{1}{s}-\tfrac{1}{t}+ H_{r,1} +\tfrac{\pi^2}{6}(s+t)    
\\& + \zeta(3) (s+t)^2  
- H_{r,1,2} st 
\\& + \tfrac{\pi^4}{360} (s+t)(4s^2 + 4t^2 +st)
\\&  -\left(\tfrac{\pi^2}{6} H_{r,2} 
{-} H_{r,1,3}\right) st(s{+}t) 
\\&  + \zeta(5) (s{+}t)^2 (s^2{+}st{+}t^2) 
\\&   +  \left[\tfrac{\pi^2}{6}H_{r,3} {-} \left(\tfrac{\pi^2}{6} {+} H_{r,2}\right)\zeta(3) {-} H_{r,1,4}\right]s t(s{+}t)^2 
\\&   + \tfrac{2r}{1+r}\;{}_{7} F_{6}\left[
\begin{subarray}{c}
1,1,1,1,1,1,1{-}r\\
2,2,2,2,2,2{+}r
\end{subarray}
;-1\right] s^2 t^2 + \cdots\, ,
}{A_EFT}
where we have defined the generalized harmonic number $H_{n,m}=\sum_{k=1}^n k^{-m}$ and its  moments $H_{n,m,k} = \sum_{\ell=1}^n H_{\ell,m}\ell^{-k}$~\footnote{Our formula for the amplitude at low energies can be extended to real $r$ via the polygamma identities $H_{r,1} = \gamma\,{+}\, \psi(1\,{+}\,r)$, $H_{r,2} = \tfrac{\pi^2}{6}\,{-}\,\psi^{(1)}(1\,{+}\,r)$, and $H_{r,3}\,{=}\,\zeta(3)\,{+}\,\tfrac{1}{2}\psi^{(2)}(1\,{+}\,r)$, along with the identities for the finite multiple harmonic sums~\cite{Yamamoto,Bala}, 
\begin{equation*}
\begin{aligned}
\hspace{6mm} H_{r,1,2} &=  \tfrac{2r}{1+r}\;{}_{5} F_{4}\left[
\begin{subarray}{c}
1,1,1,1,1{-}r\\
2,2,2,2{+}r
\end{subarray}
;-1\right] \\
H_{r,1,3} &= \tfrac{\pi^2}{6}H_{r,2} - \zeta(3)H_{r,1} + {\DOTSI\intop}_{\!\!\!0}^1 {\rm d}z\frac{1-z^r}{1-z}{\rm Li}_3(z) \\
H_{r,1,4} &= {\DOTSI\intop}_{\!\!\!0}^1 {\rm d}z \left[\tfrac{\pi^2}{6}\log z - {\rm Li}_3 (z) +\zeta(3)\right]\times\\& \qquad\qquad\times\left[z^2 \Phi(z,1,1{+}r)+z^{-1}\log(1-z)\right],
\end{aligned}
\end{equation*}
where ${\rm Li}_n$ is the polylogarithm and $\Phi$ is the Lerch transcendent.}. Note that \Eq{A_EFT} apparently exhibits uniform transcendentality~\cite{DHoker:2019blr}.

From \Eq{amplitude_hypergeo_simp} we can straightforwardly derive the behavior of the amplitude for high-energy fixed-angle scattering.  In particular, sending $|s|,|t|\rightarrow\infty$, we obtain
\eq{
A(s,t) \sim e^{B(s,t)} +\frac{r}{st}+\cdots \, ,
}{fixed_angle}
where $B(s,t) = (s+t)\log(s+t)-s\log s-t\log t +\cdots$ and ellipses denote subleading contributions we will ignore. For nonnegative integer $r$, the first and second terms in \Eq{fixed_angle} are trivially derived from the first and second terms in \Eq{amplitude_hypergeo_simp}.  In the physical region, $-1\leq \cos\theta=1+\frac{2t}{s} \leq 1$, we find that $B(s,t)<0$, so $A(s,t) \sim r/st$ falls off as a power law.  In the unphysical region, $t\;{>}\;0$, we find that $B(s,t)>0$, in which case $A(s,t)\sim e^{B(s,t)}$ exhibits the same behavior as the Veneziano amplitude, in accordance with general arguments~\cite{Caron-Huot:2016icg}. 

Meanwhile, the Regge limit of $s\rightarrow \infty$ at fixed $t$ is
\eq{
A(s,t) = s^{J(s,t)} +\frac{r}{(1+t)s} +\cdots\, ,
}{}
where $J(s,t) = t +\cdots$ and ellipses denote subleading terms.  Here the first and second terms dominate depending on whether $t>0$ or $t<0$, respectively.  Any spurious poles in $t$ will cancel between terms, as in \Eq{amplitude_hypergeo_simp}.

Rather incredibly, the hypergeometric amplitude in \Eq{amplitude_hypergeo} has an integral representation that is tantalizingly reminiscent of the string worldsheet,
\eq{
A(s,t) = r \int_0^1  \int_0^1 {\rm d}x\,  {\rm d}y \,\frac{x^{-s-1}y^{r-1} (1-xy)^t}{(1-x)^{t+1}} \, .
}{}
Amazingly, the integral above is just the Koba-Nielsen formula for the five-point Veneziano amplitude~\cite{Fairlie:1970di},
\begin{equation}
\int_0^1 \int_0^1 {\rm d}x\, {\rm d}y\, \frac{x^{{-}s_{12}{-}1}y^{{-}s_{45}{-}1}(1 -xy)^{s_{23}{+}s_{34}{-}s_{51}}}{(1 - x)^{s_{23}{+}1}(1 - y)^{s_{34}{+}1}} \, ,
\end{equation}
evaluated at $s_{12}=s$, $s_{23}=t$, $s_{34}=s_{51}=-1$, and $s_{45}=-r$. It is perhaps not so surprising that this object is a viable four-point amplitude.  By setting $s_{34}, s_{45}, s_{51}$ to constant values, we ensure that the resulting expression exhibits singularities only in $s$ and $t$, while the $s\leftrightarrow t$ symmetric choice of kinematics enforces crossing.
For the case of $r\rightarrow 0$, the five-point Veneziano amplitude factorizes onto the massless pole at $s_{45}\rightarrow 0$, so multiplying by $r$ to remove the singularity yields the four-point Veneziano amplitude, as expected.

There is an infinite parameter space of amplitudes constructed from a weighted sum over the solutions above,
\eq{
\lambda_m &= \int_{-\infty}^\infty {\rm d}r \, \mu(r)  \frac{r!}{(m+r)!}\,.
}{lambda_general}
Given any choice of $\mu(r)$ for which the integral converges, the corresponding $\lambda_m$ yield a consistent amplitude.
The resulting object is a linear combination of our hypergeometric amplitudes, so its properties are straightforward to derive.

\begin{figure*}[t]
\begin{center}
\includegraphics[width=0.8\columnwidth]{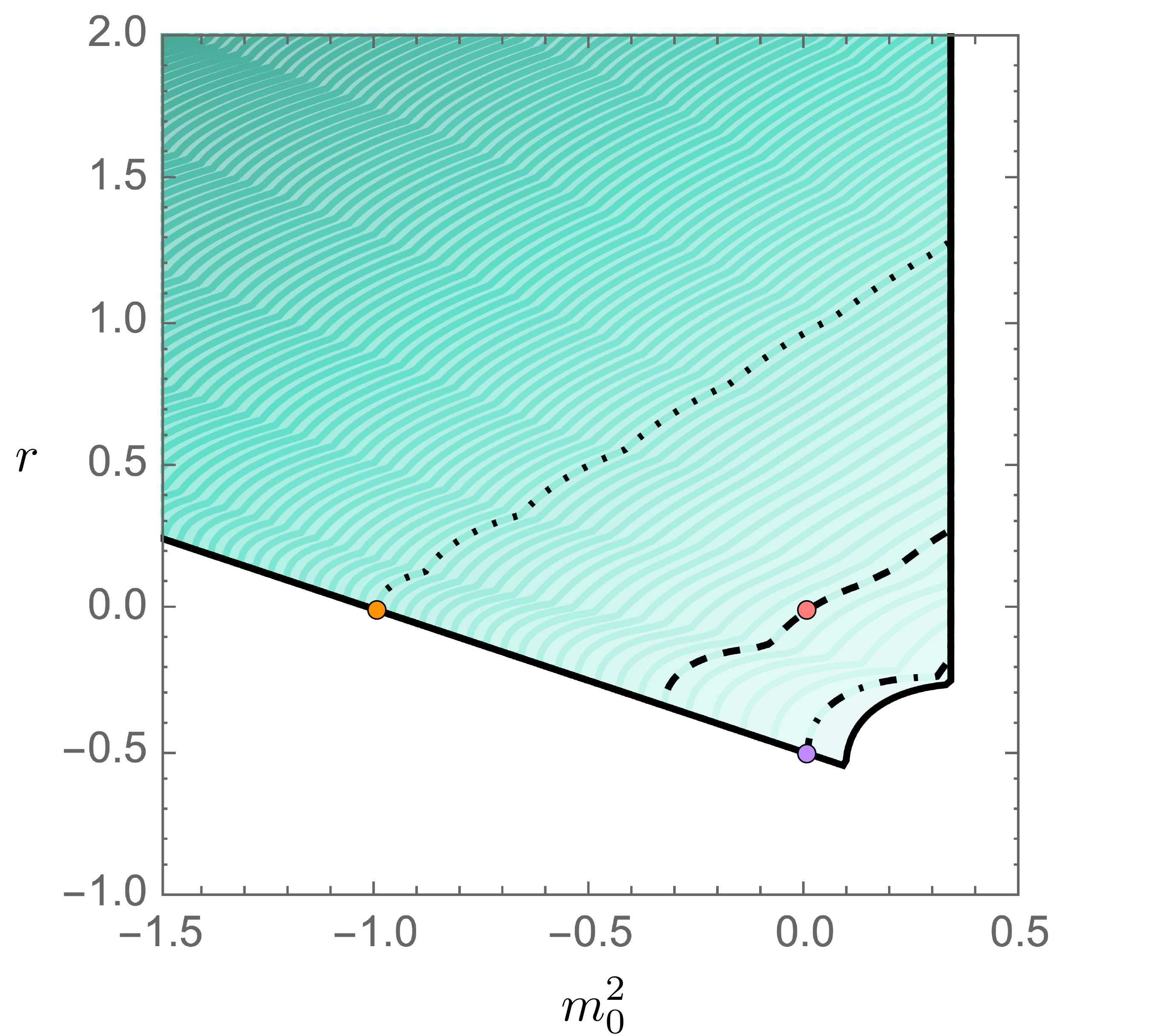} \qquad
\includegraphics[width=0.8\columnwidth]{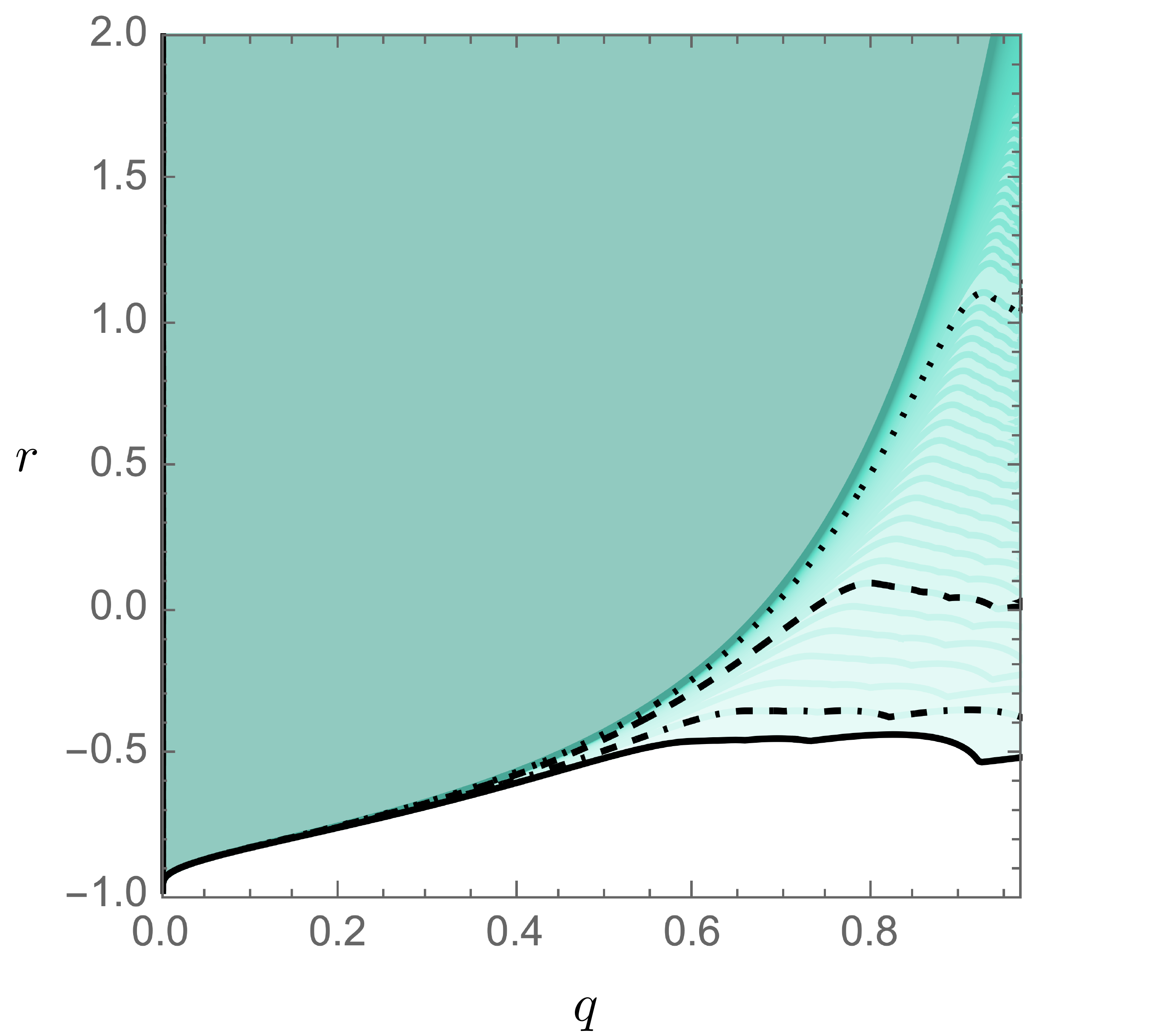}
\end{center}\vspace{-7mm}
\caption{Each line corresponds to a fixed choice of integer spacetime dimension $D\geq 4$.  The region above this line satisfies partial wave unitarity.  {\it Left panel:} $\{m_0^2, r\}$ plane at fixed $q=1$.  {\it Right panel:} $\{q,r\}$ plane at fixed $m_0^2=0$.
We have highlighted regions of physical interest, including lines at fixed $D=$ 4 (solid), 5 (dashed dotted), 10 (dashed), and 26 (dotted), together with points corresponding to the bosonic string (orange), superstring (red), and a critical case (purple) defined by the minimal value of $r$ for $m_0^2=0$, which also happens for exactly $D=5$.   The dark teal region in the right panel is unitary for all $D$.
}
\label{fig:unitarity}
\end{figure*}

\medskip

\noindent {\bf $q$-Integer Spectrum Bootstrap.}
Next, consider a spectrum defined by $q$-deformed integers,
\eq{
m^2_n = [n]_q =  \frac{1-q^n}{1-q},
}{qdefn}
where $0\,{<}\,q\,{<}\,1$~\footnote{Analyses of the $q$-integer spectrum are typically restricted to positive $q$, for the aesthetic choice of a monotonic spectrum. While $q\,{>}\,1$ fails unitarity~\cite{Figueroa:2022onw}, the case of $0\,{<}\,q\,{<}\,1$ exhibits an accumulation point, as in the case of the hydrogen atom or recent constructions in string theory~\cite{Maldacena:2022ckr}.  Monotonicity is not necessarily required, however, and an amusing example that violates this condition is $q= -\varphi^{-2}$, where $\varphi =(1+\sqrt{5})/2$ is the golden ratio.  In this case the spectrum is $[n]_{-\varphi^{-2}}= F_n /\varphi^{n-1} $, where $F_n$ is the Fibonacci sequence. }.  The $q$-deformed integers reduce to the usual integers for $q\rightarrow 1$, allowing for a simple comparison to the results of the previous section.

To streamline our analysis, we define $s=[\sigma]_q$ and $t=[\tau]_q$ in analogy with Eq.~\eqref{qdefn}, where $\sigma$ and $\tau$ are the natural kinematic variables given the spectrum.  We also assume a dual resonant representation for the amplitude,
\eq{
A(\sigma,\tau) &=\sum_{n=0}^\infty \frac{R_n([\tau]_q)}{[n-\sigma]_q} \, ,
}{amplitude_ansatz_exp}
where $R_n([\tau]_q) = R_n(t)$ is the same quantity defined in \Eq{residue_ansatz_linear}.  Furthermore, since $\frac{1}{[n-\sigma]_q} = \frac{1-(1-q)s}{[n]_q-s}$, \Eq{amplitude_ansatz_exp} exhibits simple poles in $s$ at the points $s= [n]_q$.

Next, let us consider special kinematics where $\sigma$ and $\tau$ are offset by a nonnegative integer $k$,
\eq{
\tau= \sigma-k \, .
}{}
Imposing crossing symmetry on the amplitude implies that $A(\sigma, \sigma-k) = A(\sigma-k, \sigma)$, so
\eq{
\hspace{-1mm}\sum_{n{=}k}^\infty \!\frac{R_{n}([\sigma{-}k]_q ){-}R_{n{-}k}([\sigma]_q)}{[n-\sigma]_q} \,{=}\,{-}\!\!\sum_{n{=}0}^{k-1} \frac{R_{n}([\sigma{-}k]_q)}{[n-\sigma]_q } \, .\hspace{-2mm}
}{}
As before, we are motivated to impose the crossing condition on the residues,
\eq{
R_{n}([n-k]_q ) = R_{n-k}([n]_q ) \quad \textrm{for} \quad 1\leq k\leq n \, .
}{residue_crossing_exp}
Solving these equations for the ansatz in \Eq{residue_ansatz_linear}, we obtain the general solution
\eq{
R_n([\tau]_q) 
&= \sum_{m=0}^{n}\frac{ \lambda_{m} q^{\frac{m(m-1)}{2}}}{[m]_q!}  \frac{[\tau]_q!}{[\tau\,{-}\,m]_q!}\frac{[n]_q!}{[n\,{-}\,m]_q!} \, ,
}{residue_solution_exp}
which is the $q$-deformed generalization of \Eq{residue_solution_linear}.

\medskip

\noindent {\bf Coon Amplitude.}
For the special choice of $\lambda_m = q^{m(m+1)/2}\times \tfrac{1}{[m]_q!}$,
we use the $q$-Vandermonde identity to write the residue in \Eq{residue_solution_exp} as $
R_n([\tau]_q) = \frac{[\tau +n]_q!}{[\tau]_q![n]_q!}= \begin{psmallmatrix} \tau+n \\ n \end{psmallmatrix}_q$,
which is a $q$-deformed binomial distribution. Plugging this residue back into \Eq{amplitude_ansatz_exp}, we obtain
\eq{
A(\sigma,\tau) &=  \sum_{n=0}^\infty \frac{1}{[n-\sigma]_q}  \left(
\begin{array}{c}
\tau +n \\
n
\end{array}
\right)_q \, .
}{notCoon}
Since \Eq{notCoon} is literally the $q$-deformation of the Veneziano amplitude in \Eq{eq:Veneziano}, it is natural to conjecture that it is the Coon amplitude~\cite{Coon}.   However, it is not.  In fact, the sum in \Eq{notCoon} does not even converge, so  \Eq{residue_crossing_exp} is invalid, and $s\leftrightarrow t$ crossing fails. 

However, if we dress each term in the sum by hand with an {\it additional} factor of $q^{\tau(\sigma-n)}$, we obtain 
\eq{
A(\sigma,\tau) & = \sum_{n=0}^\infty \frac{q^{\tau(\sigma-n)}}{[n-\sigma]_q}  \left(
\begin{array}{c}
\tau +n \\
n
\end{array}
\right)_q  \\ &= q^{\sigma \tau}  \frac{\Gamma_q(-\sigma)\Gamma_q(-\tau)}{\Gamma_q(-\sigma-\tau)} \, ,
}{Coon}
which is exactly the Coon amplitude~\cite{Coon,Baker:1970vxk,Coon:1972qz}.

Since the mysterious prefactor $q^{\tau(\sigma-n)}\,{=}\,1$ for $\sigma\,{=}\,n$, it does not affect the residues of the amplitude and contributes only contact interactions. Clearly,  there is an infinite space of similar functions, but only the choice made above yields the Coon amplitude.
We leave an investigation of other possible prefactors for future work.

\medskip

\noindent {\bf $q$-Hypergeometric Amplitude.} Consider a more general parameter choice
$\lambda_m\,{=}\,q^{m(m{+}1)/2{+}rm}\,{\times}\, \tfrac{ [r]_q!}{[m{+}r]_q!}$,
for which the residue is $
R_n([\tau]_q) = \frac{[\tau+n+r]_q! [r]_q!}{[\tau+r]_q! [n+r]_q!}$.
Again dressing each term in \Eq{amplitude_ansatz_exp} with $q^{\tau(\sigma-n)}$ and using the  $q$-deformed Thomae transformation~\cite{GasperRahman}, we obtain
\eq{
&A(\sigma,\tau) =  \sum_{n{=}0}^\infty \frac{q^{\tau(\sigma-n)}}{[n-\sigma]_q} \frac{[\tau+n+r]_q! [r]_q!}{[\tau+r]_q! [n+r]_q!} \\
&=q^{\sigma \tau}  \frac{\Gamma_q({-}\sigma)\Gamma_q({-}\tau)}{\Gamma_q(-\sigma-\tau)} \;
{}_{3}\phi_{2}\left[\begin{array}{c}
q^{-\sigma},q^{-\tau},q^{r}\\
q^{-\sigma-\tau} ,q^{1+r}
\end{array};q;q\right],
}{amplitude_qhypergeo}
where the last factor is a basic hypergeometric function~\footnote{The basic hypergeometric series reduces to the usual hypergeometric series in the $q\rightarrow 1^-$ limit via
\begin{equation*}
\hspace{5mm}  {}_{m}\phi_{n}\!\left[\begin{subarray}{c}
q^{\alpha_{1}},\ldots,q^{\alpha_{m}}\\
q^{\beta_{1}},\ldots,q^{\beta_{n}}
\end{subarray};\!q;\!(q{-}1)^{1{+}n{-}m}z\right]\!\!\!\stackrel{\,\,q{\rightarrow} 1^-}{=} \!\!\!{}_{m}F_{n}\!\left[\begin{subarray}{c}
\alpha_{1},\ldots,\alpha_{m}\\
\beta_{1},\ldots,\beta_{n}
\end{subarray};z\right].
\end{equation*}
\phantom{.} }.  The object in \Eq{amplitude_qhypergeo} subsumes every amplitude we have discussed thus far, including hypergeometric ($q\,{=}\,1$), Coon ($r\,{=}\,0$), and Veneziano ($\{q,r\}=\{1,0\}$). 

In analogy with \Eq{lambda_general}, we can also consider an arbitrary linear combination of the $\lambda_m$ defined above, in which case the resulting amplitude is a corresponding linear combination of $q$-hypergeometric amplitudes that subsumes all of the amplitudes in Refs.~\cite{Cremmer:1971yf,Arik:1974ed}.

\medskip

\noindent {\bf Unitarity Bounds.} There is a sizable parameter space, depicted in Fig.~\ref{fig:unitarity}, for which our new amplitudes are consistent with unitarity.
Following the analysis of Refs.~\cite{Figueroa:2022onw,Cheung:2022mkw,Bhardwaj:2022lbz,Fairlie:1994ad,Arkani-Hamed:2022gsa}, we consider external states with mass $m_0^2$, which is accomplished by simply sending $(s,t)\rightarrow (s-m_0^2,t-m_0^2)$ in the $q$-hypergeometric amplitude in \Eq{amplitude_qhypergeo}.  The resulting amplitude depends on the set of parameters $\{q,r,m_0^2\}$.

Expanding the residue of the pole at level $n$ in partial waves, we obtain $R_{n}(t) = \sum_{\ell=0}^n a_{n,\ell} G_\ell^{(D)}(\cos\theta)$, where $\cos\theta \,{=}\, 1\,{+}\,\frac{2t}{s\,{-}\,4m_0^2}$ and $G_\ell^{(D)}$ denotes the $D$-dimensional Gegenbauer polynomials.  Unitarity implies that $a_{n,\ell} \geq 0$, thus sculpting out a consistent parameter region spanned by $\{q,r,m_0^2,D\}$ \footnote{The partial waves can be analytically computed and can be expressed as sums,
\begin{widetext}
\begin{equation*}
\begin{aligned}
a_{n,\ell} &= \left(1+\tfrac{2\ell}{D-3}\right)\Gamma\left(\tfrac{D-1}{2}\right)\tfrac{(-1)^{\ell}}{(q^{-r-n};q)_{k}}{\SigmaScaled}_{j=\ell}^{n}\SigmaScaled_{s=0}^{\lfloor(j-\ell)/2\rfloor}\tfrac{j!\left[1-q^{n}+3m_0^2(q-1)\right]^{\ell+2s}\left[3-q^{n}+m_0^2(q-1)\right]^{j-\ell-2s}}{2^{\ell+2s+j}(j-\ell-2s)!s!\Gamma(\frac{D-1}{2}+\ell+s)}\tfrac{q^{r(j-n)}(q^{-n};q)_{n-j}}{(q;q)_{n-j}} 
\\&
\stackrel{q\rightarrow 1}{=} \left(1+\tfrac{2\ell}{D-3}\right)\Gamma\left(\tfrac{D-1}{2}\right)\tfrac{\Gamma\left(1+r\right)}{\Gamma\left(1+r+n\right)}\SigmaScaled_{j=\ell}^{n}\SigmaScaled_{s=0}^{\lfloor(j-\ell)/2\rfloor}S_1(n,j)\tfrac{j!(n-3m_0^2)^{\ell+2s}(2-n+2r+m_0^2)^{j-\ell-2s}}{2^{\ell+2s+j}(j-\ell-2s)!s!\Gamma\left(\frac{D-1}{2}+\ell+s\right)},
\end{aligned}
\end{equation*}
\end{widetext}
where $S_1(n,j)$ is the unsigned Stirling number of the first kind.}.  
We find numerically that $r$ is always larger than $-1$, from which we analytically find $m_0^2 \leq 1/3$ from $a_{n,n}\geq 0$. From $a_{n,n-1}\geq0$, we then have the bound $2q^{-1-r} \geq 3-q+m_0^2(q-1)$, which reduces to $r\geq -(1+m_0^2)/2$ as $q\rightarrow 1$.
We leave a full analysis of partial wave unitarity for future work.

\medskip

\noindent {\bf Gravity Bootstrap.} For the case of a gravitational amplitude $M(s,t)$ that vanishes as $s\rightarrow \infty$ at fixed $t<0$, we can compute the dispersion relation,
\eq{
M(s,t) &= \oint\limits_{s'{=}s} \frac{{\rm d}s'}{2\pi i}\frac{M(s',t)}{s'-s} =\int\limits^\infty_{-\infty} \frac{{\rm d}s'}{\pi} \frac{{\rm Im}\,M(s',t)}{s'-s} \\
 &=  \sum_{n=0}^\infty \left( \frac{1}{m^2_n-s}+\frac{1}{m^2_n-u} \right)R_n(t)\, ,
}{M_dispersion}
where we assume that the external states are massless, so $s+t+u=0$.  Hence, $M(s,t)$ can be expressed as an infinite sum over poles in the $s$ and $u$ channels.

Next, let us assume a linear spectrum as in \Eq{spectrum_linear}, so that the dual resonant form of the amplitude is
\eq{
M(s,t) &= \sum_{n=0}^\infty \left( \frac{1}{n-s}+\frac{1}{n-u} \right)R_n(t)\, ,
}{grav_dual_res}
where $R_0(t) =\xi/ t^{2}$ so that $M(s,t) = \frac{\xi}{stu}+\cdots$ exhibits a long-range force at low energies.  Since $1/stu$ is trivially crossing symmetric and dual resonant, it can be added or subtracted with impunity without violating our input assumptions.  Hence, $\xi$ is freely floating and cannot be related to any other parts of the amplitude by crossing.

For the residues at positive $n$, we define
\eq{
R_n(t) =\sum\limits_{m=0}^{2(n-1)}  \kappa_{n,m} t^m\, ,
}{grav_ansatz}
so the modes exchanged at level $n$ carry up to spin $2(n\,{-}\,1)$.  For graviton scattering, $M(s,t)$ is dressed with a polarization-dependent prefactor ${\cal R}^4$ that carries spin weight, in which case these modes carry up to spin $2n$.

Next, we restrict to $t=s-k$, which implies that $u=-2s+k$.   Since \Eq{grav_dual_res} is $s\leftrightarrow u$ symmetric, we need only enforce crossing on $s \leftrightarrow t$.  Equating $M(s,s-k)$ to $M(s-k,s)$ implies that $\sum_{n=1}^\infty \frac{R_n(s-k)}{n-s}+\frac{R_n(s-k)}{n-k+2s} =  \sum_{n=1}^\infty \frac{R_n(s)}{n+k-s}+\frac{R_n(s)}{n-k+2s}$.
Relabeling the summation and equating the residue of each $s$-channel pole at $s=n$, we obtain \Eq{residue_crossing_linear}.  Meanwhile, equating the residue of each $u$-channel pole at $s=-\tfrac{n-k}{2}$, we obtain
\eq{
R_n(-\tfrac{n+k}{2}) =R_n(-\tfrac{n-k}{2}) \,,
}{residue_crossing_linear_perm}
where $1\,{\leq}\,k\,{\leq}\,n$. 
Next, we constrain the ansatz in \Eq{grav_ansatz} with Eqs.~\eqref{residue_crossing_linear} and \eqref{residue_crossing_linear_perm}, modulo any constraints involving $\xi$, since $1/stu$ can be freely added or subtracted from the amplitude as discussed above.  We thus  eliminate $\kappa_{n,m}$ for $m<2n$ in terms of the unfixed parameters $\kappa_m \equiv \kappa_{m,2m}$, yielding
\eq{
R_n(t)\,
{=} \sum_{m{=}0}^{n}\frac{ \kappa_{m}}{m!}  \frac{(t{+}n{+}m)!}{(t{+}n)!}\frac{(t{-}1)!}{(t{-}m{-}1)!} \frac{(n{-}1)!}{(n{-}m{-}1)!}  \, ,
}{residue_solution_linear_grav}
where $\kappa_m$ must be chosen so that the sum converges.  For a $q$-integer spectrum, the above procedure yields  no solutions, in accordance with general arguments \cite{Geiser:2022exp}.

\medskip

\noindent {\bf Virasoro-Shapiro Amplitude.} Next, consider the parameter choice
$\kappa_m = \tfrac{1}{(m+1+r)!(m+1-r)!}$. In this case, the residue becomes $R_n(t) =   \tfrac{(t+n-1+r)!}{(t+r)! (n+r)!} \tfrac{(t+n-1-r)!}{(t-r)! (n-r)!}$,
which inserted back into the amplitude in \Eq{grav_dual_res} yields
\eq{
\hspace{-2mm}M(s,t) 
\,{=}\, \frac{\Gamma(-s) }{t^2{-}r^2}   {}_4  \tilde F_3\left[
\begin{array}{c}
1,-s,t+r,t-r\\
1-s,1+r,1-r
\end{array}
;1\right] {+}  s{\leftrightarrow}u\,.\hspace{-2mm}
}{M_VS_r}
By numerical evaluation, one can verify that \Eq{M_VS_r} is not crossing symmetric for generic $r$.  This is possible because the constraints in Eqs.~\eqref{residue_crossing_linear} and \eqref{residue_crossing_linear_perm} are not sufficient conditions for crossing.  That said, for any integer $r$, \Eq{M_VS_r} is in fact crossing symmetric, yielding a generalization of the Virasoro-Shapiro amplitude,
\eq{
M(s,t) 
\,{=}\, \frac{(-1)^{r+1}\Gamma(-s\,{+}\,r)\Gamma(-t\,{+}\,r)\Gamma(-u\,{+}\,r)}{\Gamma(1\,{+}\,s\,{+}\,r)\Gamma(1\,{+}\,t\,{+}\,r)\Gamma(1\,{+}\,u\,{+}\,r)} \, ,
}{M_VS_r_int}
which in fact forms a basis for the amplitudes described in the conclusions of Ref.~\cite{Arkani-Hamed:2020blm}.

Last but not least, we can actually take the principle of dual resonance even further, writing the Virasoro-Shapiro amplitude as a sum over $s$-channel poles alone,
\eq{
M(s,t) &= \sum_{n=0}^\infty \frac{R_n(s,t)}{n-s} \, ,
}{}
where
$R_n(s,t) = \frac{(t+s-1)!}{t!s!}\frac{(t+n-1)!}{t!n!}$
depends on both $s$ and $t$, and the sum is convergent for $t<0$.

\medskip

\noindent {\bf Future Directions.}  The present work offers many lines of inquiry for future study. First and foremost, it would be interesting to bootstrap new amplitudes with different mass spectra.   Second, there is the important question of whether our new amplitudes generalize to higher-point scattering.   Finally, it would be interesting to perform a systematic analysis of the unitary regions of parameter space for our amplitudes.

\vspace{3mm}

\noindent {\it Acknowledgments:}  We thank Zohar Komargodski, Julio Parra-Martinez, John Schwarz, and Sasha Zhiboedov for comments.
 C.C. is supported by the Department of Energy (Grant No.~DE-SC0011632) and by the Walter Burke Institute for Theoretical Physics.
G.N.R. is supported at the Kavli Institute for Theoretical Physics by the Simons Foundation
(Grant No.~216179) and the National Science Foundation (Grant No.~NSF PHY-1748958)
and at the University of California, Santa Barbara by the Fundamental Physics Fellowship.

\vspace{-5mm}

\bibliographystyle{utphys-modified}
\bibliography{string_bootstrap}

\end{document}